\documentclass[11pt, executivepaper]{article}
\usepackage[utf8]{inputenc}
\usepackage[T1]{fontenc}
\usepackage{natbib}

\usepackage{tikz}
\usetikzlibrary{calc,positioning,arrows,arrows.meta,decorations.markings}

\usepackage{amsmath, amsthm, amssymb}
\usepackage{amsfonts}
\usepackage{MnSymbol}
\usepackage{mathtools}
\usepackage{xcolor}
\usepackage{graphicx}
\usepackage{enumitem}
\usepackage{geometry}
\usepackage{hyperref}
\hypersetup{colorlinks= true, allcolors=blue}
\setcitestyle{aysep={}}

\newtheorem{prop}{Proposition}

\begin{document}

\title{\textbf{On the Common Logical Structure of Classical and Quantum Mechanics}}

\author{Andrea Oldofredi\thanks{Centre of Philosophy, University of Lisbon, Portugal. E-mail: aoldofredi@letras.ulisboa.pt} \and Gabriele Carcassi\thanks{Physics Department, University of Michigan, Ann Arbor, MI 48109. E-mail: carcassi@umich.edu} \and Christine A. Aidala\thanks{Physics Department, University of Michigan, Ann Arbor, MI 48109. E-mail: caidala@umich.edu}}

\maketitle

\begin{abstract}
At the onset of quantum mechanics, it was argued that the new theory would entail a rejection of classical logic. The main arguments to support this claim come from the non-commutativity of quantum observables, which allegedly would generate a non-distributive lattice of propositions, and from quantum superpositions, which would entail new rules for quantum disjunctions. While the quantum logic program is not as popular as it once was, a crucial question remains unsettled: what is the relationship between the logical structures of classical and quantum mechanics? In this essay we answer this question by showing that the original arguments promoting quantum logic contain serious flaws, and that quantum theory does satisfy the classical distributivity law once the full meaning of quantum propositions is properly taken into account. Moreover, we show that quantum mechanics can generate a distributive lattice of propositions, which, unlike the one of quantum logic, includes statements about expectation values which are of undoubtable physical interest. Lastly, we show that the lattice of statistical propositions in classical mechanics follows the same structure, yielding an analogue non-commutative sublattice of classical propositions. This fact entails that the purported difference between classical and quantum logic stems from a misconstructed parallel between the two theories.
\vspace{4mm}

\end{abstract}
\vspace{5mm}

\begin{center}
\emph{Accepted for publication in Erkenntnis}
\end{center}
\clearpage

\tableofcontents
\vspace{5mm}

\section{Introduction}

As soon as Quantum Mechanics (QM) achieved a definite and coherent mathematical formulation, several physicists and philosophers claimed that its formal structure does not conform to the laws of classical propositional calculus.\footnote{For an introduction and a clear discussion of quantum logic the reader may refer to cf.\ \cite{Giuntini:2002}, \cite{DallaChiara:2004}, \cite{deRonde:2016}, \cite{Wilce:2021}.} The first modifications of Classical Logic (CL) were advanced in the thirties in order to faithfully represent the physical content of quantum theory. For instance, in 1931 the Polish philosopher Zygmunt Zawirski proposed to apply \L ukasiewicz's three-valued logic to QM, starting from considerations about the wave-particle duality and Heisenberg's uncertainty principle. In virtue of the latter, moreover, the logical empiricists Moritz Schlick and Philipp Frank claimed that the conjunction of two or more statements may have no meaning in quantum theory, concluding that the classical conjunction is not always valid in the context of quantum physics, as recalled in \cite{Carnap:1966}. In their view, a typical example of invalid conjunction is one attributing a definite value for both position and momentum to a single quantum particle at a given time $t$. Furthermore, in 1933 Fritz Zwicky suggested that QM rejects the law of excluded-middle.\footnote{For historical details on quantum logic see \cite{Jammer:1974}.} However, it is generally recognized that the founding text of Quantum Logic (QL) is Birkhoff and von Neumann's essay ``\emph{The Logic of Quantum Mechanics}'' (\cite{vonNeumann:1936}), where the authors provided the standard account of the propositional calculus obeyed by quantum propositions.\footnote{It is worth noting that von Neumann in his treatise \emph{Mathematische Grundlagen der Quantenmechanik} published in 1932 anticipated that from the algebraic structure of quantum theory it would have been possible to formulate a new propositional calculus. These ideas would be fully expressed in his successive collaboration with Garrett Birkhoff.}

A few years after the publication of this paper, the research in quantum logic saw a notable development in both the scientific and philosophical communities, opening new research programs searching for the correct logical structures to understand and describe the physics of QM (cf.\ \cite{Reichenbach:1944}, \cite{Mackey:1957}, \cite{Finkelstein:1963}, \cite{Kochen:1965}, \cite{Jauch:1969}, \cite{Giuntini:2002}, \cite{DallaChiara:2004}, \cite{Pitowsky:2006}, \cite{Giuntini:2018}, \cite{Svozil:2020}, \cite{Holik:2021}), and intense metaphysical debates regarding the nature of logic itself and its empirical status (cf.\ notably \cite{Quine:1951}, \cite{Putnam:1968}, \cite{Dummett:1976}, \cite{Hallett:1982}, \cite{Weingartner:2004}, \cite{Bacciagaluppi:2009}). 

The belief according to which QM entails a new logic stood the test of time, and it is still alive to the present day. For example, \cite{Giuntini:2018} claim that ``[t]he \emph{logic} of classical physical objects is naturally based on a two-valued semantics. Such a dichotomic situation breaks down in quantum theory'' (p.\ 4), implying that the latter entails a revision of classical propositional calculus. Such a view is shared by \cite{Wilce:2021}, who says: ``[m]athematically, quantum mechanics can be regarded as a non-classical probability calculus resting upon a non-classical propositional logic''. Similarly, explaining the difference between the logic of classical mechanics and the logic obeyed by quantum propositions, \cite{Holik:2019} write that
\begin{quote}
[i]n classical and quantum mechanics, the physical properties of a system are endowed with a lattice structure. These structures are different in the classical
and quantum case, and they determine the logical structure of the physical system (\cite{Holik:2019}, p.\ 363-364).
\end{quote}
\noindent More precisely the authors say:
\begin{quote}
The distributive inequalities are the main difference between classical and quantum logic. In the classical lattice, all properties satisfy the distributive equalities, but in the quantum lattice, only distributive inequalities hold, in general (\emph{ibid.}).
\end{quote}

Even more explicitly, presenting the standard formulation of quantum theory and its peculiar features, Stefanovich states that
\begin{quote}
the true logical relationships between results of measurements are different from the classical laws of Aristotle and Boole. The usual classical logic needs to be replaced by the so-called quantum logic (\cite{Stefanovich:2019}, p.\ 2).
\end{quote}

While the quantum logic programme is still being actively pursued by a sizable community, and despite the many interesting applications in the fields of quantum information, computation and cryptography that are being investigated to the present day, QL is no longer viewed as a solution to the foundational problems affecting quantum theory.\footnote{This conclusion is now accepted among experts; for details see e.g.\ \cite{Bacciagaluppi:2009} and \cite{Giuntini:2002}. Referring to this, the latter authors stated explicitly that ``quantum logics are not to be regarded as a kind of ``clue'', capable of solving the main physical and epistemological difficulties of QT [quantum theory]. This was perhaps an illusion of some pioneering workers in quantum logic'' (\cite{Giuntini:2002}, p.\ 225).} Nevertheless, although it may be reasonable for a physicist to simply discard the approach over practical considerations, such as the lack of a desired result, on philosophical grounds this move would beg important questions, as for instance: what is the relationship between the logical structures of classical and quantum mechanics? How are they different and how are they similar? What should we think about the core insights that led to the development of QL? Were they substantially correct, or should they be rejected? We think that providing a correct answer to these issues will lead (i) to a deeper understanding of the structural relations existing between classical logic and quantum mechanics---which unfortunately are not yet precisely understood---and (ii) to a clarification of important confusions concerning the motivations behind QL. %For these reasons, the present essay is focused on the logical structures common to both quantum and classical mechanics, highlighting the similarities between these two theoretical frameworks.
 \footnote{In the present essay we will be concerned only with propositional calculus in the context of non-relativistic QM. First-order and higher-order logics will not be discussed here, nor the logic of relativistic quantum theories.}

Referring to this, one of us recently faced the question of whether quantum physics implies a revision of classical logic (cf.\ \cite{Oldofredi:2020}). Alternatively stated, it was asked if classical logic must be inevitably abandoned in the quantum realm, or if there are quantum theories compatible with classical propositional calculus. Analyzing Bohmian mechanics, a well-known interpretation of QM, it has been shown that quantum physics does not necessarily involve a departure from classical logic. More precisely, it has been argued that in this interpretation of QM not only the logical connectives retain their classical meanings---in particular the negation and the disjunction do follow classical rules---but also the distributivity law holds. This fact is a consequence of the particle ontology of the theory and its explanation of quantum measurements.

In this paper we will be concerned with a similar issue, since we ask whether standard quantum mechanics itself necessarily involves a new logical structure radically different with respect to classical propositional calculus. In what follows we will answer this question in the negative.\footnote{It is worth noting that in what follows we are \emph{not} going to discuss whether it is useful or possible to describe quantum phenomena with a non-classical type of logic.} More precisely, this essay has a two-fold aim: in the first place, we analyze the main motivations and justifications for the introduction of quantum logic, claiming that such arguments do not really entail the failure of CL in the quantum context because they contain serious flaws. In the second place, we show that the mathematical structure on which QM is built already provides us with a distributive lattice of quantum propositions, contrary to the received view in QL. Such a distributive lattice of quantum statements, moreover, allows us to take into account assertions concerning expectation values, which are generally excluded in the standard approach to quantum logic but are nonetheless experimentally relevant. Finally, we argue that classical probabilistic theory is also equipped with an analogue non-distributive lattice. All these aspects, to our knowledge, are not sufficiently discussed in literature; thus, the present paper can help to shed new light on the relation between logic and quantum mechanics.
\vspace{2mm}

The essay is structured as follows: in Section \ref{QL} we review the main arguments that have been advanced in the literature to introduce a quantum logical calculus; here we also provide a brief outline of the standard approach to QL (readers familiar with Birkhoff and von Neumann's proposal can skip this part). Section \ref{Motivations} explains why the main examples used to show the failure of classical logic in QM contain important flaws and ambiguities; moreover, we argue that the classical distributive law can be maintained in quantum mechanics, once such ambiguities are removed. In Section \ref{Math}, contrary to the received view in QL, we show that quantum theory already possesses a distributive lattice of propositions in the form of the Borel algebra, which includes not only the lattice of standard quantum logic, but also important sentences such as those concerning expectations values. We also show that classical probabilistic theory possesses a formally analogue structure. In Section \ref{conc} we discuss the main philosophical implications of the preceding sections and conclude the paper.

\section{The Standard Approach to Quantum Logic}
\label{QL}

Given that a significant part of the arguments in favor of QL stem from a claimed unsuitability of classical logic in the context of quantum mechanics, let us say a few words about classical logic, classical mechanics and their purported relationship.  It is a well known result that classical propositional logic is equivalent to a Boolean algebra where propositions (atomic and complex) have truth values ``true'' and ``false'' (or $\top$ and $\bot$), and the main operations among them are conjunction ``$\wedge$'', disjunction ``$\vee$'' and negation ``$\neg$'', the only unary operation. From these logical connectives one can introduce secondary operations, as for instance the material implication ``$\longrightarrow$'', the exclusive or ``$\oplus$'' and logical equivalence ``$\longleftrightarrow$''. Furthermore, it is important for our discussion to underline that the laws of propositional logic include commutativity, associativity, identity and distributivity with respect to $\wedge$ and $\vee$. It is also well established that the algebraic structure underlying classical mechanics is commutative, distributive and associative. The standard view in QL is that classical logic is applicable to classical mechanics because of its commutative algebraic structure.

The idea is that the logical operations of conjunction, disjunction and negation correspond in the context of classical mechanics respectively to multiplication, addition and complementation among the observables of the theory. This has an interesting physical significance, since these observables associated to magnitudes of physical systems can be added and multiplied together---i.e.\ one can sum and multiply measurement results: ``since the observables [of CM] form a commutative Poisson algebra, addition and multiplication of observables reflect the action of adding and multiplying results of different measurements'' (cf.\ \cite{David:2015}, p.\ 77). Both propositional logic and classical mechanics, thus, generate a complemented Boolean lattice; from this fact it follows that the observable algebra of CM ``is isomorphic to a Boolean algebra of propositions with $(\wedge, \vee, \neg)$'' (\cite{David:2015}, p. 79).
To this regard, Beltrametti claims that
\begin{quote}
Boolean algebras are algebraic models of classical logic (more specifically of classical propositional calculus) with the algebraic operations of meet ($\wedge$) and join ($\vee$) corresponding to the logical connectives ``and'', ``or'', and the unary relation of orthocomplementation corresponding to the logical negation. The rules and tautologies of classical logic have their counterpart in equations and identities of Boolean algebras (\cite{Beltrametti:2004}, p.\ 341). 
\end{quote}

Therefore, it is possible to formally characterize the state of a physical system via logical propositions which can assume the truth values ``true'' or ``false''---depending whether such statements describe a true or false state of affairs concerning the system under consideration (cf.\ \cite{Jaeger:2009}, p. 61)---and to perform logical operations among propositions via the connectives $(\wedge, \vee, \neg)$. It is worth noting that in the context of CM it is always determined whether a system instantiates a given property or not; alternatively stated, every logical proposition about physical systems is either true or false, and hence in CM the principle of semantic bivalence holds (for more details cf.\ \cite{Bub:2007}, p.\ 642, \cite{Giuntini:2002}, p.\ 130).

Once the previous stance is taken, one necessarily concludes that classical propositional calculus is not appropriate to represent the logic obeyed by quantum propositions, since quantum theory does not share the same algebraic structure of classical mechanics. Referring to this, \cite{Giuntini:2002} claim that distributivity, a fundamental property of classical logic, fails in quantum logic for two main reasons:
\begin{itemize}
	\item the non-commutativity of quantum observables,
	\item the peculiar behavior of the quantum disjunction---another major deviation from classical logic---which may be true even though neither of the members is true in virtue of the linearity of the Schr\"odinger Equation (SE). 
\end{itemize}
\noindent Let us then have a closer look at these motivations for the failure of the distributivity law, and more generally of CL, in the quantum domain.

Contrary to the case of classical mechanics, QM relies on a non-commutative algebraic structure\footnote{For details on the mathematical structure of QM and its physical content the reader may refer e.g.\ to \cite{Sakurai1994}, and \cite{Griffiths:2014}. In this essay, it is assumed that the reader has some familiarity with the standard formalism of QM.}: quantum observables generally do not commute, meaning that for any pair of operators $A,B$ we have $[A,B]=AB-BA\neq0$. From a physical perspective this fact entails that if one performs a measurement of $A$ on a quantum system followed by a measurement of $B$, in general one will obtain different results inverting the order of these observations. Moreover, in virtue of the non-commutative algebra of quantum theory, measuring an operator $C=A+B$ is not typically equivalent to measuring $A$ and $B$ independently and then adding the respective results, since $A,B$ will be in general incompatible (cf.\ \cite{David:2015}, p.\ 77). This feature of QM constitutes the formal basis to prove the Heisenberg uncertainty relation, a theorem of quantum theory which reflects the operational inability to simultaneously prepare/measure the values of incompatible operators with arbitrary precision. Such a result, in turn, is usually interpreted ontologically in the sense that quantum systems do not instantiate definite properties in non-measurement situations (cf.\ \cite{Sakurai1994}).

In addition, as we will explain in the remainder of this section, the failure of distributivity in quantum logic is due to another particular characteristic of this new propositional calculus, namely the behavior of the quantum disjunction, which is different w.r.t.\ its classical counterpart in virtue of the presence of quantum superpositions. Indeed, another remarkable difference between QM  and CM is that quantum systems can be in a superposition of states as a consequence of the linearity of the Schr\"odinger equation---the fundamental dynamical law of quantum theory---which for a single particle reads: 
\begin{align}
	\label{SE}
	i\hbar\frac{\partial\psi}{\partial t}=\Big(-\frac{\hbar^2}{2m}\nabla_k^2+V\Big)\psi=H\psi,
\end{align}
\noindent where $H$ represents the Hamiltonian operator, defined as the sum of kinetic and potential energy of the system at hand. More specifically, this algebraic property of \eqref{SE} entails that if two wave functions\footnote{In QM the wave function of a system provides the maximal information available about it.} $\psi_1, \psi_2$ are both possible solutions of the same Sch\"odinger equation, then their linear combination (\emph{superposition}) 
\begin{align}
	%\label{SSE}
	\psi_s=\alpha\psi_{1}+\beta\psi_{2} \nonumber
\end{align} 
is still a solution of the same SE---$|\alpha|^2,|\beta|^2$ (with $\alpha,\beta\in\mathbb{C}$) represent the probabilities to find the system in $\psi_1, \psi_2$ respectively. Notably, the new superposed state $\psi_s$ is also a consistent representation of the system. Referring to this, \cite{deRonde:2016} explicitly say that in QL 
\begin{quote}
	differently from the case in classical semantics, a quantum disjunction may be true even if neither of its members is true. This reflects, for example, the case in which we are dealing with a state such as that of a spin $1/2$ system which is in a linear combination of states up and down.
\end{quote}
\noindent Let us use this exact case to review both alleged problems, the peculiar nature of the quantum disjunction and the lack of distributivity in quantum logic, with a simple example taken from \cite{Giuntini:2002}.

Let us consider a spin-$1/2$ particle, such as an electron. From QM we know that there are only two possible spin states in which a particle can be found along each axis after a spin measurement, namely either in the up or in the down state. Furthermore, since spin operators along different axes are incompatible and do not commute---i.e.\ $[S_x, S_y]\neq 0, [S_x, S_z]\neq 0, [S_y, S_z]\neq0$---they represent incompatible observables that cannot be simultaneously measured. Let us then consider the following scenario: an electron is prepared with spin up along the $x$ direction and subsequently the spin along the $y$ axis is measured. Let the propositions $p$, $q$ and $r$ be defined as follows:
\begin{description}
	\item $p$: ``the electron has $x$-spin up'', 
	\item $q$: ``the electron has $y$-spin up'',
	\item $r$: ``the electron has $y$-spin down''.
\end{description}
As we said, there are only two states in which a particle can be found along a given axis. Therefore the proposition $q \vee r$ ``must be true'' (\cite{Giuntini:2002}, pp.\ 133-134). However, because the different directions are incompatible, if $p$ is true then both $q$ and $r$ must be false. The result is, as noted before, that the quantum disjunction is true even if neither of its members is true.

This behavior in turn breaks the law of distributivity which would state:
\begin{align}
	\label{DL}
	p\wedge(q\vee r)\longleftrightarrow(p\wedge q)\vee(p\wedge r).
\end{align}
\noindent Since we assume that $p$ is true and we know that the disjunction $(q\vee r)$ is also true, we deduce that $p\wedge(q\vee r)=\top$; however, $(p\wedge q)$ and $(p\wedge r)$ are both false in virtue of the incompatibility between the spin operators among different axes. This fact implies that the right hand side of the distributive law asserts a false statement, i.e.\ $(p\wedge q)\vee(p\wedge r)=\bot$. Consequently, we have 
\begin{align}
	\label{contradiction}
	p\wedge(q\vee r)=\top\longleftrightarrow(p\wedge q)\vee(p\wedge r)=\bot. 
\end{align}
Therefore the distributive law fails in the context of quantum theory.\footnote{In the context of quantum logic, it is usually believed that distributivity must be replaced by a weaker law: $(p\wedge q)\vee(p\wedge r)\longrightarrow(p\wedge(q\vee r)).$}

These types of problems were highlighted by Birkhoff and von Neumann, who took them as evidence for the failure of classical logic in the context of quantum theory. The principal aim of their essay on quantum logic was to ``discover what logical structures one may hope to find in physical theories which, like quantum mechanics, do not conform to classical logic'' (\cite{vonNeumann:1936}, p.\ 823). The expression \emph{quantum logic} in this paper must be clarified, since it refers to a quantum propositional calculus in the form of a ``calculus of linear subspaces with respect to \emph{set products}, \emph{linear sums}, and \emph{orthogonal complements}'' which ``resembles the usual calculus of propositions with respect to \emph{and}, \emph{or}, and \emph{not}'' (\emph{ibid}.), where the logical propositions are associated to measurements, tests on quantum systems.\footnote{Cf.\ \cite{Giuntini:2002} and \cite{DallaChiara:2004} for a systematic introduction to various forms of quantum logic, and to \cite{Engesser:2009} for historical and philosophical discussions on the topic.}

Their analysis begins by defining quantum logical propositions as \emph{experimental propositions}, i.e.\ statements affirming that a certain observable $A$ (or a set of observables) measured on a quantum system $s$ has a given value $a_i$ (or a sequence of values). More precisely, the authors stress that both in classical and quantum mechanics observations of physical systems are given by the readings of the experimental outcomes ($x_1, \dots, x_n$) of \emph{compatible} measurements ($\mu_1, \dots, \mu_n$). The values ($x_1, \dots, x_n$) are elements of what the authors called the ($x_1, \dots, x_n$)-space, i.e.\ the ``observation-space'' of the system in question, whose elements are all the possible combinations of results of the compatible measurements ($\mu_1, \dots, \mu_n$). Hence, the actual values ($x_1, \dots, x_n$) form a subset of such a space. Birkhoff and von Neumann, then, defined the experimental propositions concerning a physical system as the subsets of the observation space associated with it.

Secondly, the authors underlined that in classical and quantum mechanics the states of physical systems are mathematically represented by points in their state spaces---phase space for the classical case, and Hilbert space $\mathcal{H}$ for the quantum case---which provide the maximal information concerning the system. A point in phase space corresponds to the specification of the position and momentum variables of a certain classical system, whereas in QM the points of $\mathcal{H}$ correspond to wave functions. The authors, then, find a connection between subsets of the observation-space of a system and subsets of its Hilbert space, specifying that quantum experimental propositions are mathematically represented by a closed linear subspace of $\mathcal{H}$; this step is crucial in order to obtain the quantum propositional calculus.\ Alternatively, we can say that quantum mechanical operators correspond to propositions with ``yes/no'' (``true/false'') outcomes in a logical system\footnote{Cf.\ also \cite{David:2015}, p.\ 78, where we read that ``An orthogonal projector $\textbf{P}$ onto a linear subspace $P\subset\mathcal{H}$ is indeed the operator associated to an observable that can take only the values 1 (and always 1 if the state $\psi\in P$ is in the subspace $P$) or 0 (and always 0 if the state $\psi\in P^{\perp}$ belongs to the orthogonal subspace to $P$).\ Thus we can consider that measuring the observable $\textbf{P}$ is equivalent to perform a test on the system, or to check the validity of a logical proposition $p$ on the system''.} as underlined by Svozil:
\begin{quote}
Any closed linear subspace of---or, equivalently, any projection operator on---a Hilbert space corresponds to an elementary
proposition. The elementary \emph{true--false} proposition can in English be spelled out explicitly as
\begin{quote}
``The physical system has a property corresponding to the associated closed linear subspace'' (\cite{Svozil:1999}, p.\ 1).
\end{quote}
\end{quote}

Thus, since in quantum mechanics physical systems do not have well-defined values for their properties before the measurement of a given operator (contrary to their classical counterparts), or better that quantum observations do not reveal pre-existing values, we have to stress that such quantum propositions refer always to measurements or to preparation procedures. Hence, the sentence ``the particle $s$ has $x$-spin up'' means either that we have measured the observable $S_x$ and we have found the value $+\hbar/2$ associated with the state ``$x$-spin-up'', or equivalently that we prepared $s$ in the $x$-spin-up state. Thus, one cannot simply say, as in classical mechanics, that a certain system has a certain property without having prepared such a system in a given state or having measured it.\footnote{More on this below.}

Now that we have clarified what quantum propositions are, let us introduce Birkhoff and von Neumann's quantum logic. In order to properly define a propositional calculus for QM, one must define the logical operators for conjunction, disjunction and negation and the notion of logical implication. Following \cite{vonNeumann:1936}, the procedure is rather simple: 
\begin{enumerate}
\item The negation of a proposition $p$ is defined by the authors as follows: ``since all operators of quantum mechanics are Hermitian, the mathematical representative of the negative of any experimental proposition is the orthogonal complement of the mathematical representative of the proposition itself'' (\cite{vonNeumann:1936}, pp.\ 826-827). The orthogonal complement $P^{\perp}$ of the subspace $P$ is the set whose elements are all the vectors orthogonal to the elements of $P$. Such an orthogonal complement satisfies the following property: given a subset $P\subset\mathcal{H}$ and a pure state $\psi$, $\psi(P)=1$ iff $\psi(P^{\perp})=0$ and $\psi(P)=0$ iff $\psi(P^{\perp})=1$. As Dalla Chiara and Giuntini underline, ``$\psi$ assigns to an event probability 1 (0, respectively) iff $\psi$ assigns to the orthocomplement of $P$ [notation adapted] probability 0 (1, respectively). As a consequence, one is dealing with an operation that \emph{inverts} the two extreme probability-values, which naturally correspond to the truth-values \emph{truth} and \emph{falsity} (similarly to the classical truth-table of negation)'' (\cite{Giuntini:2002}, p.\ 132).
\item Concerning the conjunction, Birkhoff and von Neumann note that one can retain the very same set-theoretical interpretation of the classical conjunction in QL, since the intersection of two closed subspaces $P, Q\subset\mathcal{H}$ is still a closed subspace. Thus, one maintains the usual meaning for the conjunction: the pure state $\psi$ verifies $P\cap Q$ iff $\psi$ verifies both $P$ and $Q$. Thus, quantum logic does not introduce a new logical operator for the conjunction.
\item Contrary to the previous case, the logical operator for disjunction cannot be represented by the set-theoretic union, since the set-theoretic union of the two subspaces $P, Q$ will not be in general a subspace; thus, it is not an experimental proposition. Therefore, in QL one introduces the quantum logical disjunction as the the closed span of the subspaces $P, Q$, which is an experimental proposition. Such a statement corresponds to the ``smallest closed subspace containing both $P$ and $Q$'' (\cite{Bacciagaluppi:2009}, p.\ 55). 
\item As far as logical implication is concerned, \cite{vonNeumann:1936} claim that given two experimental propositions $p$ and $q$ about a certain physical system, ``$p$ implies $q$'' means that ``whenever one can predict $p$ with certainty, one can predict also $q$ with certainty'' (p.\ 827) and this is equivalent to stating that ``the mathematical representative of $p$ is a subset of the mathematical representative of $q$'' (\emph{ibid.}). This fact is particularly important since the authors showed that ``it is \emph{algebraically} reasonable to try to correlate physical qualities with subsets of phase-space'' and thus, ``physical qualities attributable to any physical system form a partially ordered system'' (\cite{vonNeumann:1936}, p.\ 828, notation adapted).
\end{enumerate}

We have thus reviewed the standard arguments to support the failure of classical logic in the context of quantum mechanics, and we have reviewed the basic mathematical structure used in quantum logic.

\section{Quantum Disjunction and Distributivity Law}
\label{Motivations}

In this section we will review the examples given in the previous one and explain that the failure of distributivity stems from the ambiguity of the quantum logical propositions and, once that is clarified, standard classical rules can be applied also in the quantum context. More precisely, we are going to argue that (i) it is not necessarily the case that a quantum disjunction can be true when neither of its members is\footnote{Clearly, a quantum disjunction is false when both disjuncts are false; however, the interesting case for our discussion is the one in which quantum propositions have undetermined truth values.}---i.e.\ when both disjuncts have undetermined truth values as in the case discussed above in which we considered a system in a linear combination of $y$-spin states---and (ii) that \eqref{DL} can be retained in QM, since the arguments seen above---usually meant to show its failure---are flawed.

\subsection{The Quantum Disjunction}

As noted in the previous section, in classical mechanics one generally considers physical systems as objects having well-defined values for inherent, dynamical and relational properties which do not depend on any observation---meaning that measuring a magnitude on a classical particle will reveal a pre-existing value of the observed quantity. Thus, statements about the ontic state of classical systems before or after a measurement can be conflated without any dangerous metaphysical consequence.\footnote{Alternatively stated, in classical mechanics it is usually claimed that the sentence ``The value of the quantity $y$ for the system $s$ is $x$'' can be considered equivalent to the statement ``After a measurement $M$ of the quantity $y$ on the system $s$, the value $x$ is obtained''. This equivalence is due to the fact that measurements in the classical context do not alter the observed system, or in less idealized scenarios, that the disturbance caused by the interaction between observed system and measuring device is negligible. A more elaborate treatment of measurements in classical mechanics goes beyond the scope of the present paper.}

However, in the context of QM this move is not allowed: the effect of a measurement on the state of a quantum system, however one may choose to interpret it, cannot be disregarded. A statement about a system before the performance of a certain measurement---i.e.\ about its preparation\footnote{In operational quantum theory the preparations are a set of instructions that an agent has to follow in order to prepare a quantum system in a certain state. We are aware that preparations can be considered special kinds of measurements. For the sake of the argument, however, in what follows we take preparations to be the instructions to arrange a physical system in a given state \emph{before} the measurement of a certain quantity is actually performed.}---is different from a statement regarding the same object when the measurement has been actually carried out, i.e.\ about the obtained measurement outcome. Indeed, it is only through repeated observations on identically prepared systems that we can infer anything about the state of the system before the measurement. Thus, propositions like ``the electron has $y$-spin up'', ``the velocity of the particle is $v$'', ``the position of the particle is $x$'' are inherently ambiguous as they do not spell out whether they are concerned with preparations or measurement outcomes. Once such vagueness is taken into account and clarified, the purported failure of classical logic disappears.

In order to support this claim, let us redefine the propositions encountered in the previous example as follows:
\begin{description}
	\item $p_i$: ``before the measurement, the electron $x$-spin is up'', 
	\item $p_o$: ``after the measurement, the electron $x$-spin is up'', 
	\item $q_i$: ``before the measurement, the electron $y$-spin is up'',
	\item $q_o$: ``after the measurement, the electron $y$-spin is up'',
	\item $r_i$: ``before the measurement, the electron $y$-spin is down'',
	\item $r_o$: ``after the measurement, the electron $y$-spin is down''.
\end{description}
\noindent Given the setup of the physical situation at hand, after the measurement of $S_y$ either $q_o$ or $r_o$ must be true, which means $q_o \vee r_o = \top$. This reflects the simple fact that \emph{after} a measurement of spin along the $y$ axis the state of the quantum particle must be one of the two indicated in either $q_o$ or $r_o$. Nonetheless, it is worth noting that $q_i$ and $r_i$ are \emph{not} the same statements as $q_o$ and $r_o$---i.e.\ $q_i$ and $r_i$ are not equivalent to $q_o$ and $r_o$---therefore, it is not necessarily true that $q_o \vee r_o = q_i \vee r_i$. This unwarranted equivalence is what causes the problem. To make evident why it is so, let us add a third step to our example. 

After the measurement of $S_y$, in the case we obtained $y$-spin down we will rotate the direction of spin by 90 degrees using the precession generated by a magnetic field, so that $x$-spin will be up, as we can see from Figure 1 below. However, in the other case, when $y$-spin up is obtained, we leave the direction of spin unchanged.

\begin{figure}
\tikzstyle{vecArrow} = [thick, decoration={markings,mark=at position
	1 with {\arrow[semithick]{open triangle 60}}},
double distance=1.4pt, shorten >= 5.5pt,
preaction = {decorate},
postaction = {draw,line width=1.4pt, white,shorten >= 4.5pt}]
\tikzstyle{innerWhite} = [semithick, white,line width=1.4pt, shorten >= 4.5pt]

\begin{center}
\scalebox{0.9}{
	\begin{tikzpicture}
		\coordinate (o) at (-3,0);
		\draw[-{Stealth[scale=1.8]}] ($ (o) + (0, -3) $) -- ($ (o) + (0, 3) $);
		\node at ($ (o) + (.25, 2.5) $) {$y$};
		\draw[-{Stealth[scale=1.5]}] ($ (o) + (-.985, -.3) $) -- ($ (o) + (.985, .3) $);
		\node at ($ (o) + (.7, -.2) $) {$x$};
		
		\coordinate (a) at (0,0);
		\draw[dashed] ($ (a) + (0, 0) $) ellipse (1 and 3);
		\draw[-{Stealth[scale=1.8]}, dashed] ($ (a) + (0, 0) $) -- ($ (a) + (0, 3) $);
		\node at ($ (a) + (-.2, 1.3) $) {$q_i$};
		\draw[-{Stealth[scale=1.5]}, thick] ($ (a) + (0, 0) $) -- ($ (a) + (.985, .3) $);
		\node at ($ (a) + (.6, -.1) $) {$p_i$};
		\draw[-{Stealth[scale=1.8]}, dashed] ($ (a) + (0, 0) $) -- ($ (a) + (0, -3) $);
		\node at ($ (a) + (-.2, -1.3) $) {$r_i$};
		\coordinate (aT) at ($ (a) + (0, -3.5) $);
		\node[align=center] at (aT) {\small Step 1 \\ Preparation};
		
		\coordinate (bb) at (1.5,0);
		\draw[vecArrow] ($ (bb) + (-0.2, 0) $) to ($ (bb) + (0.2, 0) $);
		\draw[innerWhite] ($ (bb) + (-0.2, 0) $) to ($ (bb) + (0.2, 0) $);
		
		\coordinate (bc) at (3, 0);
		\draw ($ (bc) + (-1.2, -.6) $) rectangle ($ (bc) + (+1.2, .6) $);
		\node at ($ (bc) $) {Measurement};
		
		\coordinate (bd) at (4.5,0);
		\draw[vecArrow] ($ (bd) + (-0.2, 0) $) to ($ (bd) + (0.2, 0) $);
		\draw[innerWhite] ($ (bd) + (-0.2, 0) $) to ($ (bd) + (0.2, 0) $);
		
		\coordinate (b) at (6,0);
		\draw[dashed] ($ (b) + (0, 0) $) ellipse (1 and 3);
		\draw[-{Stealth[scale=1.5]}, thick] ($ (b) + (0, 0) $) -- ($ (b) + (0, 3) $);
		\node at ($ (b) + (-.2, 1.3) $) {$q_o$};
		\draw[-{Stealth[scale=1.8]}, dashed] ($ (b) + (0, 0) $) -- ($ (b) + (.985, .3) $);
		\node at ($ (b) + (.6, -.1) $) {$p_o$};
		\draw[-{Stealth[scale=1.5]}, thick] ($ (b) + (0, 0) $) -- ($ (b) + (0, -3) $);
		\node at ($ (b) + (-.2, -1.3) $) {$r_o$};
		\coordinate (bT) at ($ (b) + (0, -3.5) $);
		\node[align=center] at (bT) {\small Step 2 \\ Measurement};
		
		\coordinate (cb) at (7.5,0);
		\draw[vecArrow] ($ (cb) + (-0.2, 1.5) $) to ($ (cb) + (0.2, 1.5) $);
		\draw[innerWhite] ($ (cb) + (-0.2, 1.5) $) to ($ (cb) + (0.2, 1.5) $);
		\draw[vecArrow] ($ (cb) + (-0.2, -1.5) $) to ($ (cb) + (0.2, -1.5) $);
		\draw[innerWhite] ($ (cb) + (-0.2, -1.5) $) to ($ (cb) + (0.2, -1.5) $);
		
		\coordinate (cc) at (9, 0);
		\draw[dashed] ($ (cc) + (-1.2, -.6) + (0, 1.5) $) rectangle ($ (cc) + (+1.2, .6) + (0, 1.5) $);
		\node at ($ (cc) + (0, 1.5) $) {No rotation};
		\draw ($ (cc) + (-1.2, -.6) + (0, -1.5) $) rectangle ($ (cc) + (+1.2, .6) + (0, -1.5) $);
		\node at ($ (cc) + (0, -1.5) $) {90$^o$ rotation};
		
		\coordinate (cd) at (10.5,0);
		\draw[vecArrow] ($ (cd) + (-0.2, 1.5) $) to ($ (cd) + (0.2, 1.5) $);
		\draw[innerWhite] ($ (cd) + (-0.2, 1.5) $) to ($ (cd) + (0.2, 1.5) $);
		\draw[vecArrow] ($ (cd) + (-0.2, -1.5) $) to ($ (cd) + (0.2, -1.5) $);
		\draw[innerWhite] ($ (cd) + (-0.2, -1.5) $) to ($ (cd) + (0.2, -1.5) $);
		
		\coordinate (c) at (12,0);
		\draw[dashed] ($ (c) + (0, 0) $) ellipse (1 and 3);
		\draw[-{Stealth[scale=1.5]}, thick] ($ (c) + (0, 0) $) -- ($ (c) + (0, 3) $);
		\node at ($ (c) + (-.2, 1.3) $) {$q_f$};
		\draw[-{Stealth[scale=1.5]}, thick] ($ (c) + (0, 0) $) -- ($ (c) + (.985, .3) $);
		\node at ($ (c) + (.6, -.1) $) {$p_f$};
		\draw[-{Stealth[scale=1.8]}, dashed] ($ (c) + (0, 0) $) -- ($ (c) + (0, -3) $);
		\node at ($ (c) + (-.2, -1.3) $) {$r_f$};
		\coordinate (cT) at ($ (c) + (0, -3.5) $);
		\node[align=center] at (cT) {\small Step 3 \\ Rotation};
		
		%	\draw \setA;
		%	\begin{scope}
		%		\clip \setA;
		%		\draw (-5, 4) .. controls (0, 2) and (0, -2) .. (-5, -4);
		%		\draw (2, 4) .. controls (2, 2) and (0, 1) .. (-1.46, 1);
		%		\draw (1, 1.9) .. controls (1.5, 1) and (2, -2.5) .. (0.5, -4);
		%		\draw (1.49, -1)  .. controls (2, -0.3) and (4, 0.5) ..  (5, 0);
		%	\end{scope}
		%	\node at (-4, -4.5) {$A$};
		%	\node at (-3.2, 0) {$P_1$};
		%	\node at (-0.4, 2.6) {$P_2$};
		%	\node at (-0.2, -1.2) {$P_3$};
		%	\node at (3, -1.7) {$P_4$};
		%	\node at (3, 1.5) {$P_5$};
           \end{tikzpicture} 
}
\end{center}
\caption{\small{Graphic representation of the discussed example. Step 1 describes the preparation of the electron in the $x$-spin up state; Step 2 represents the measurement of $S_y$ at a later time; finally in Step 3 we rotate the direction of spin by 90 degrees if at Step 2 the particle is found in the $y$-spin down state. We leave the direction of spin unchanged otherwise.}}
\end{figure}
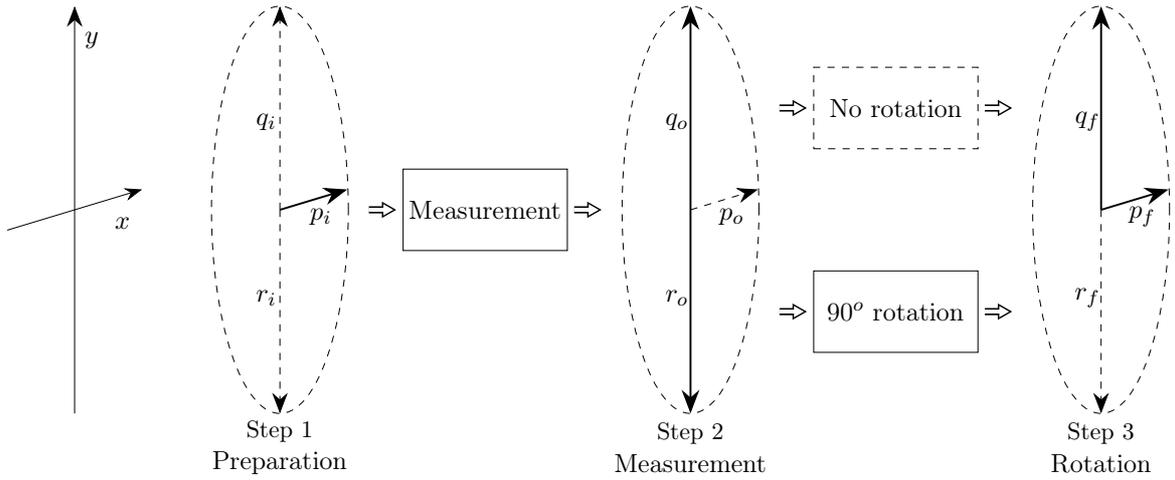

Consider now the statements:
\begin{description}
	\item $p_f$: ``after the third step, the electron $x$-spin is up'',
	\item $q_f$: ``after the third step, the electron $y$-spin is up'',
	\item $r_f$: ``after the third step, the electron $y$-spin is down'',
\end{description}
\noindent that concern the state of the electron at the later time, after the possible rotation of spin. By construction, the particle will either have $x$-spin up or $y$-spin up, either $p_f$ or $q_f$ must be true and therefore $p_f \vee q_f = \top$.\footnote{In fact, note that $q_f$ is true if and only if $q_o$ is true, and, similarly, $p_f$ is true if and only if $r_o$ is true.} But again, this relationship is valid only after the third step, which means that $p_o \vee q_o$ is not necessarily true because these latter statements are evaluated at different (earlier) times. 

This situation is formally identical to the previous one: before we had that $q_o \vee r_o \neq q_i \vee r_i$, now we have $p_f \vee q_f \neq p_o \vee q_o$. It would be improper to say that an electron has always either $x$-spin up or $y$-spin up just because it is so after the third step. Similarly, it would be improper to claim that the electron has always either $y$-spin up or down just because it is so after the measurement of spin along the $y$ axis. Neither the disjunction, nor its atomic components tell us anything about an inherent property associated with the observable $S_y$ possessed by a quantum particle independently of any measurement of $S_y$.

On the other hand, we can calculate the truth value of $q_i \vee r_i$ directly and show that it is false. The difficulty here is that before the measurement of $S_y$ only $x$-spin is defined, so how are we going to compute statements about the value of $y$-spin? The solution is simple: while the value of an observable is not always defined, its expectation value always is. The idea is then to translate $q_i$ and $r_i$ into propositions about expectation values, which we already know how to handle, and apply the standard rules. Consider the following statements:
\begin{description}
	\item $q'_i$: ``before the measurement, the expectation value for $y$-spin is $\frac{1}{2} \hbar$'',
	\item $r'_i$: ``before the measurement, the expectation value for $y$-spin is $-\frac{1}{2} \hbar$''.
\end{description}
\noindent If $\rho_i$ is the (potentially) mixed state before the measurement, the expectation value for $y$-spin is given by $\text{Tr}(\rho_i S_y) = \frac{1}{2} \hbar \langle \psi_y^+ | \rho_i | \psi_y^+ \rangle - \frac{1}{2} \hbar \langle \psi_y^- | \rho_i | \psi_y^- \rangle$. Therefore $q'_i$ is true if and only if $\text{Tr}(\rho_i S_y) = \frac{1}{2} \hbar$. Because $\rho_i$ is positive definite, this can only happen if $\langle \psi_y^+ | \rho_i | \psi_y^+ \rangle = 1$ and $\langle \psi_y^- | \rho_i | \psi_y^- \rangle = 0$, which means $\rho_i = | \psi_y^+ \rangle \langle \psi_y^+ |$. This entails that the expectation value for $y$-spin is $\frac{1}{2} \hbar$ if and only if the electron was prepared in a pure state of $y$-spin up. In other words, because the spin of the electron along any axis is bounded by $\pm \frac{1}{2} \hbar$, the only way that $q'_i$ is true is if $q_i$ is true. We have that $q'_i$ and $r'_i$ are respectively equivalent to $q_i$ and $r_i$. Consequently $q_i \vee r_i = q'_i \vee r'_i$. Note that $q'_i \vee r'_i$ is the proposition ``before the measurement, the expectation value for $y$-spin is $\pm \frac{1}{2} \hbar$''. Now, before the measurement of $S_y$ we know that the electron $x$-spin is up. While the $y$-spin value is not fully defined---the state is not an eigenstate of $S_y$---the expectation value is fully defined and it is zero. Zero is different from $\pm \frac{1}{2} \hbar$ and therefore $q_i \vee r_i = \bot$. In fact, since any (non-trivial) superposition of $y$-spin up and down will result in an expectation value strictly between $-\frac{1}{2}\hbar$ and $\frac{1}{2}\hbar$, $q_i \vee r_i$ will be true if and only if either $q_i$ or $r_i$ is true. Therefore it is not the case that the quantum disjunction can be true even if none of the elements is true. The overall point is that the incompatibility of the observables does indeed introduce a problem because it is not clear how to evaluate propositions that mix different variables. The use of equivalent propositions over expectation values helps us solve the ambiguities.

While a full discussion goes beyond the scope of this work, we should mention that it is rather odd to claim that one of the reasons QM fails to follow classical logic is the linearity of the SE given that linear systems abound in classical mechanics as well. All wave equations are linear which include both the SE and classical electromagnetism. The Liouville equation, the classical analogue to the SE that governs the evolution of probability distributions under classical Hamiltonian mechanics, is also linear. One can create superpositions of classical electromagnetic fields or classical probability distributions in the same way one does of wave functions, and this does not seem to imply an invalidation of classical logic.

\subsection{Keeping Distributivity in Quantum Mechanics}

Several authors have already argued in different ways that it is possible to retain distributivity in quantum mechanics. For instance, \cite{Park:1968} disputed the identification between Hermitian operators and quantum observables and provided a new interpretation of quantum measurement theory rejecting the notion of incompatible observables---and thereby incompatible measurements---which, as we have seen, constitutes the basis to show the failure of the distributivity law in the quantum realm. Remarkably, also Bohr argued that classical logic should be maintained in quantum theory as a consequence of the complementarity principle:
\begin{quote}
The aim of the idea of complementarity was to allow of keeping the usual logical forms while procuring the extension necessary for including the new situation relative to the problem of observation in atomic physics (Bohr quoted in \cite{Faye:2021}, p.\ 115).
\end{quote}
Another interesting quote reported in \cite{Faye:2021} clearly shows that the Danish physicist strongly opposed the idea of replacing classical propositional calculus with QL: 
\begin{quote}
The question has been raised whether recourse to multivalued logics is needed for a more appropriate representation of the situation. From the preceding argumentation it will appear, however, that all departures from common language and ordinary logic are entirely avoided by reserving the word `phenomenon' solely for reference to unambiguously communicable information, in the account of which the word `measurement' is used in its plain meaning of standardized comparison (Bohr quoted in \cite{Faye:2021}, p.\ 115).
\end{quote}

Although Bohr did not provide formal arguments against the introduction of quantum logic, his views express the idea according to which macroscopic measurement results must be described with ordinary language, therefore, they must be subjected to the rules of classical logic---in his view the intrinsic novelties of quantum mechanics have to be found in the contextual and complementary nature of quantum phenomena on the one hand, and the entanglement between quantum systems and macroscopic devices which in measurement situations form an indissoluble unity on the other. Hence, given that quantum propositions concern only measurement results, Bohr concluded that they should be governed by the laws of classical propositional calculus.

To this regard, \cite{Heelan:1970} provided a formalization of Bohr's arguments based on the complementarity principle and showed that the statements about a single quantum mechanical event, i.e.\ a measurement of a certain observable given a precise experimental context, do follow a classical propositional calculus.\footnote{We thank one anonymous reviewer for having pointed out Heelan's work to us.} More precisely, Heelan claims that QM introduces a distinction between events, corresponding to the actual performance of a certain measurement, and physical contexts, i.e.\ experimental settings determining the necessary and sufficient conditions for the realization of a particular measurement outcome. In his view, then, one has to introduce two different languages: an event-language useful to formally describe a particular observation of a certain quantum observable and a context-language, which is a meta-language necessary in order to properly speak about how a certain event can possibly occur.\footnote{Heelan exemplifies such a distinction as follows: ``If the event, for example, is a particle-location event, the event-language is position-language, and the physical context is a standardized instrumental set-up plus whatever other physical conditions are necessary and sufficient for the measurement of a given range of possible particle-position events'' (\cite{Heelan:1970}, p.\ 96).} If we consider only propositions involving and relating incompatible contexts, Heelan says, then we will generate a lattice of propositions which is not distributive, however, if we take into account single event-languages---each of which is coupled with a given context-language---the logic of such individual quantum events can be classical. This captures Bohr's idea according to which the principle of complementarity arises from the context-dependent character of quantum mechanical events.

Although we partially agree with Heelan's view in saying that propositions about individual quantum measurements can generate a distributive lattice, we want to generalize such a claim. In fact, we show---with simpler arguments and independently of Bohr's interpretation of quantum theory---that taking into account the temporal order of individual quantum measurements one can relate even statements about incompatible observations into a distributive lattice. In addition, we show that ignoring the temporal order of measurements may lead to violations of the distributivity law also in classical mechanics; thus, it is our aim to argue that once the correct interpretation of quantum propositions is taken into account the lattice generated by such statements is distributive.

In order to derive our desired result, let us now turn our attention to distributivity and try to rewrite \eqref{contradiction} by using the refined propositions that explicitly include the temporal dependence. The proposition $p$ always referred to preparation, therefore it maps to $p_i$ in all cases. The disjunction was argued to be true on the basis of what happened \emph{after} the measurement of $S_y$, therefore on the left side of the relationship we should use $q_o$ and $r_o$. On the right side, instead, it was argued that $p \wedge q$ was false because of the incompatibility between the spin operators among different axis. This means we are considering statements at equal time with $p_i$, and therefore we should use $q_i$ and $r_i$. We have 
\begin{align}
	\label{solved_contradiction}
	p_i \wedge (q_o \vee r_o) = \top \longleftrightarrow (p_i \wedge q_i) \vee (p_i\wedge r_i) = \bot.
\end{align}
\noindent But this is hardly a failure of distributivity, since the statements on the left side are not the same as the statements on the right side. Let us fix that.

Before we saw that if $p_i$ is true, the expectation for $y$-spin is zero, and therefore $q'_i \vee r'_i = q_i \vee r_i = \bot$. Therefore
\begin{align}
	\label{solved_distrib_1}
	p_i \wedge (q_i \vee r_i) = \bot \longleftrightarrow (p_i \wedge q_i) \vee (p_i\wedge r_i) = \bot.
\end{align}
\noindent On the other hand, consider $p_i \wedge q_o$ which states that we find $y$-spin up after the measurement and $x$-spin up before. This is not false, and in fact it will be true in 50\% of the cases according to the Born rule. Similarly, $p_i \wedge r_o$ will be true in the remaining 50\%. Therefore we have
\begin{align}
	\label{solved_distrib_2}
	p_i \wedge (q_o \vee r_o) = \top \longleftrightarrow (p_i \wedge q_o) \vee (p_i\wedge r_o) = \top.
\end{align}
\noindent In either case, the distributivity law is satisfied.

To sum up, \eqref{DL} fails in the quantum domain because we considered propositions evaluated at different times, before and after the measurement, in the two sides of the biconditional. If we remove the ambiguity and make the temporal dependence explicit, the distributivity law is recovered. The lack of commutativity of the observables in quantum mechanics therefore does not change the rules of logic. It imposes that the lattice of propositions at different times depends on the process, in particular on the choice of measurement. That is, if at time $t$ a particular observable is either prepared or measured, all statements about incompatible observables will be impossible and therefore they will be false. Quantum measurements do have an impact on the lattice of statements, unlike classical measurements, but that impact does not change the rules of logic.

The idea that the same statement may or may not be possible at different moments in time is not particular to quantum mechanics. We can, in fact, construct a scenario in classical mechanics where propositions have the same before/after logical relationships. Suppose we have a ball sitting at position $p$ over a hatch (see Figure 2). Once the hatch opens, the ball drops, bounces around, until it rests at either position $q$ or $r$ with equal chance.

\begin{center}
\includegraphics[scale=.8]{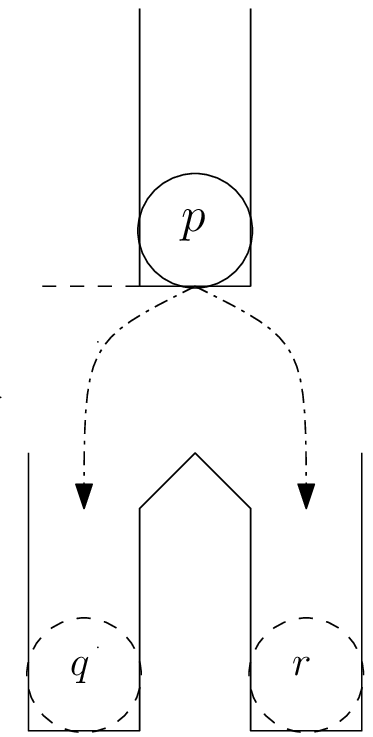}
\begin{quote}
\small{Figure 2: Graphic representation of the discussed example.  Initially a ball is in position $p$ over a hatch.  After the hatch opens, the ball drops and falls to either position $q$ or $r$ with equal chance.}
\end{quote}
\end{center}

Let us consider the following statements:
\begin{description}
    \item $p_i$: ``Before the hatch is opened, the position of the ball is $p$'',
    \item $q_i$: ``Before the hatch is opened, the position of the ball is $q$'',
    \item $r_i$: ``Before the hatch is opened, the position of the ball is $r$''.
\end{description}
\noindent It is straightforward to see that these are three incompatible assertions: since the ball can have only one possible determinate position in space before the hatch is opened, it cannot be localized in two different locations, $p$ and $q$ or $p$ and $r$. Therefore the statements obey \eqref{solved_distrib_1}. After the hatch is opened, the ball will either settle in $q$ or $r$. Consider, then, the sentences:
\begin{description}
	\item $p_o$: ``After the hatch is opened, the position of the ball is $p$'',
	\item $q_o$: ``After the hatch is opened, the position of the ball is $q$'',
	\item $r_o$: ``After the hatch is opened, the position of the ball is $r$''.
\end{description}
Since the ball must be in one of the two lower positions after it dropped, the disjunction $(q_o \vee r_o) \equiv \top$. Therefore the statements obey \eqref{solved_distrib_2}. To no surprise, the distributive law is satisfied and therefore classical logic applies.

We can, however, evaluate $q$ and $r$ at a different time w.r.t.\ $p$, as it is done in the quantum scenario considered in Section \ref{QL}. In that case, we obtain the same relationship as in \eqref{solved_contradiction}. If we omit the indices, that is, we are vague about the temporal evaluation, we get the same failure of distributivity as in \eqref{contradiction}. Naturally, one would not say that distributivity fails in the classical context, but rather that we have to evaluate propositions taking into account the same temporal instant on both sides of \eqref{DL}.

In conclusion, contrary to the received view in quantum logic, we have shown that the distributivity law is not violated in the quantum context once the correct temporal order of quantum propositions---and consequently their actual meaning---is considered.

\section{On the Common Logical Structure between Quantum and Classical Mechanics}
\label{Math}

After having critically assessed the general motivations to introduce quantum logic as a consequence of the physical content of QM, in this section we present our main argument according to which it is possible to assign a distributive lattice to quantum propositions. Remarkably, in what follows we will see that the non-distributivity of the quantum logic lattice originates neither from non-commuting observables nor from superpositions of quantum mechanics: it is a property of the lattice of subspaces of \emph{any} vector space, and of the lattice of subgroups of any group more in general (cf.\ \cite{Davey:2002}). In this regard, indeed, we also show that taking into account the space of classical probability distributions one can apply a non-distributive lattice to classical propositions. This fact is due to the similar structures of the state space of quantum mechanics and the space of classical probability distributions. 

In what follows, then, it will be argued that both theories---quantum and classical---may be assigned \emph{two} lattices of propositions: a distributive one that follows the laws of classical logic, and a non-distributive one obeying the rules of quantum logic. Contrary to the widespread opinion, therefore, we will claim not only that quantum physics does not necessarily entail a revision of classical propositional calculus, but also that it is actually possible to show that quantum propositions can form a distributive lattice. Furthermore, in agreement with the example of the violation of the distributivity law in the classical context seen above, in the remainder of the section we will argue also that the propositions of classical mechanics may obey a non-classical type of logic. 

Before we proceed, it is worth noting that the classical space we will be using is not phase-space itself, but the space of probability distributions over phase-space, which is a real Hilbert space and thus has a similar structure to its quantum counterpart. It should be stressed that probabilistic theories are important even in classical mechanics. Due to the nature of experimental verification, also in the context of classical mechanics there are always going to be uncertainties in the preparation and measurements, and repeated observations will directly lead to statistical constructs.
However, we want to emphasize that this paper aims primarily at showing that QM can generate a distributive lattice of propositions. Here we want to enter neither into debates concerning the ontology of quantum mechanics, nor into discussions about the similarities between the commitments of QM and those of classical mechanics. In this essay we simply point out a formal similarity between the above mentioned mathematical structures that has implications for the logic of both theories. 

From a mathematical perspective, the distributive lattice will be provided in both the classical and the quantum case by a Borel algebra, which is the smallest $\sigma$-algebra on a set $X$ equipped with a topological structure that contains the topology; the non-distributive lattice will be provided by the set of closed subgroups ordered by inclusion. We will claim that the non-distributive lattice is a strict subset of the distributive one; this result entails that everything that can be described in quantum logic can also be described in classical logic. Moreover, the distributive lattice will include propositions concerning the expectation value of observables that are of high scientific interest, since in an experimental context by ``measurement'' one typically refers to the statistical collection of many ``single takes''. On the contrary, it is well-known that standard quantum logic deals only with single-take measurements, remaining silent w.r.t.\ propositions involving statistical statements. Hence, measurements of cross sections, the most typical type of measurement in the context of particle physics, are not part of the quantum lattice, and neither are measurements of the probabilities themselves. Therefore, showing that one may apply a distributive lattice to quantum propositions can be seen as an important result in order to extend the set of meaningful logical propositions of quantum mechanics.  

\subsection{The Distributive Lattice of Quantum Mechanics}

As anticipated above, in this section our arguments are fundamentally based on the notion of $\sigma$-algebra, which is a significant structure from a foundational perspective, given that physically relevant mathematical branches like measure theory (cf.\ \cite{Cohn:2013}) and probability theory (cf.\ \cite{Grimmett:2001}) are built on top of it. Let us then review some of its essential features.

In order to begin our discussion, let us consider in the first place a set $X$: in the quantum case it will correspond to the set of all possible quantum states; similarly, in the classical case it will correspond to the set of all possible probability distributions over phase space.\footnote{We should stress, however, that $\sigma$-algebras are defined independently of what the elements of $X$ actually are. For a simpler example, $X$  could correspond to the set of the real numbers for a continuous quantity.} By definition, a $\sigma$-algebra is a collection of subsets of $X$ closed under countable union, countable intersection and complement with respect to $X$; the closure condition requires that  performing those operations on the sets contained in a $\sigma$-algebra will return another set in the same $\sigma$-algebra. Interestingly, these sets can be understood to represent statements concerning the objects described by $X$, i.e.\ quantum states and classical probability distributions.\footnote{In the quantum case this is exemplified also in Section \ref{QL}.} Namely, if $U \subseteq X$, the statement ``$x$ is in $U$'' tells us that the state $x$ must be one of those delimited by $U$. Then, by virtue of the equivalence between sets and statements, it will be more convenient for us to conceive $\sigma$-algebras as collections of propositions about $X$ instead of sets. Referring to this, since the $\sigma$-algebra is closed under countable union, countable intersection and complement with respect to $X$, the corresponding statements will be closed under countable conjunction, countable disjunction and negation. Hence, the $\sigma$-algebra generates a Boolean algebra of statements.\footnote{Note that the interpretation used here is exactly the same used in probability theory. In fact, a probability space is defined by three elements: a sample space $\Omega$ which represents the set of all possible outcomes; a $\sigma$-algebra over $\Omega$ which represents all events, all propositions; and a measure that assigns a probability to each event.} 

In the second place, it is crucial to underline that every topological space---as for instance the complex Hilbert space of quantum mechanics and the real Hilbert space of classical probability distributions considered in this essay---comes already equipped with a $\sigma$-algebra called a Borel algebra.\ A topology, in fact, is also a collection of sets---called \emph{open sets}---and the Borel algebra is defined to be the smallest $\sigma$-algebra that contains all the open sets.\footnote{Here the topology is important for us only as a means to construct the Borel algebra.} Complete normed spaces, such as the mentioned Hilbert spaces, have a canonical way to construct a topology and, therefore, a $\sigma$-algebra. Let us see how this procedure works in the easier case of a standard three-dimensional Euclidean space. Given two points $x$ and $y$, we can write the square of distance $\epsilon^2 = |y - x|^2$ as the square of the norm of the vector between the two points. The statement ``the position is within $\epsilon$ meters of $x$'' corresponds to the set $U = \{y \; | \; |y - x|^2 < \epsilon^2 \}$ of all the points whose distance from $x$ is less than $\epsilon$. These are exactly the open sets that generate the topology, and the Borel algebra will be the smallest $\sigma$-algebra that contains these sets.\footnote{We can now see precisely that by generating a lattice of statements we mean starting with a set of propositions that are experimentally accessible (i.e.\ the basis of the topology) and closing under negation (i.e.\ complement) and countable disjunction (i.e.\ union).} This works analogously for all normed spaces. A Hilbert space, in particular, comes with an inner product which also defines a norm. We can define a distance $\epsilon^2 = |\psi - \phi|^2=\langle \psi - \phi , \psi - \phi \rangle$ between two states, and the statement ``the state of the system is within $\epsilon$ of $\psi$'' corresponds to the set $U = \{\phi \; | \; |\psi - \phi|^2 < \epsilon^2\}$. As we have already stated, the space of classical probability distributions is also a Hilbert space where each point corresponds to the square root of the density.\footnote{The choice of the square root of the density rather than the density itself is merely a technical choice.} In this case we have a distance $\epsilon^2 = |\sqrt{\rho_1} - \sqrt{\rho_2}|^2=\langle \sqrt{\rho_1} - \sqrt{\rho_2} , \sqrt{\rho_1} - \sqrt{\rho_2} \rangle$ and we can proceed in the same exact way. In all normed spaces, the topology, and therefore the $\sigma$-algebra, is constructed from sets of the form ``the object is within $\epsilon$ of $x$''.

It is a simple result that the Borel algebra thus constructed contains all propositions of quantum logic, and we state it in the following proposition:

\begin{prop}
	Let $\mathcal{H}$ be a Hilbert space. Let $\Sigma$ be the Borel $\sigma$-algebra induced by its inner product. Let $L(\mathcal{H})$ be the lattice of closed subspaces of $\mathcal{H}$. Then $L(\mathcal{H}) \subset \Sigma$.
\end{prop}

Proof. Let $p \in L(\mathcal{H})$ be a closed sub-space of $\mathcal{H}$. Then $p$ is a closed set, the complement of an open set, and therefore a Borel set: $p \in \Sigma$. Let $v \in \mathcal{H}$ be a vector in the Hilbert space and consider the singleton $U = \{ v \}$. This is not a closed sub-space and therefore $U \notin L(\mathcal{H})$. As the topology of a Hilbert space is Hausdorff, all singletons are closed and therefore $U \in \Sigma$.\footnote{To give more context, the set $U = \{ v \}$ can be written as $U = \{w \; | \; |w - v|^2 = 0\}$ the set of all elements with zero distance from $v$. This can be understood as the limit (i.e.\ countable intersection) of a sequence of propositions ``the object is within $\epsilon$ of $v$'' where $\epsilon$ tends to zero.} This means $L(\mathcal{H}) \subset \Sigma$. This concludes the proof.

The above proposition contains our first crucial conclusion. As we stated in Section \ref{QL}, in quantum logic the propositions correspond to closed subspaces, which are closed sets and, thus, are Borel sets. All the propositions of quantum logic are consequently readily found in the Borel algebra of the Hilbert space. The Borel algebra, in turn, is a Boolean algebra that follows classical logic, and contains all the statements of quantum logic.\footnote{Obviously, the inclusion map from the lattice of closed subspaces is not an order isomorphism as the join and complement are not preserved. However, these are still expressible as subspace closure and subspace complement.} This construction also applies identically for the classical cases, where we will also be able to talk about the set of all closed subspaces.

While all of this may seem suspect to some readers, we would like to stress that these are very standard constructions. Alternatively stated, all of what has been mentioned so far are the standard mathematical techniques used to study these spaces and can be found in any textbook (cf.\ \cite{Rudin:1991}, \cite{Vasudeva:2017}). Here we are bringing out the full physical significance of these mathematical constructs, which is too often overlooked. Understanding the link between the mathematical structures and the physics they represent is crucial to clearing out possible misunderstandings.

Now that we have briefly seen the definitions and basic constructions, we can gain more insights into these structures and what they represent. First, let us understand why the lattice of subspaces (i.e. quantum logic) fails to be a Boolean algebra. Stone's representation theorem tells us that any Boolean algebra can be expressed as an algebra of sets using the standard set operations (i.e.\ intersection, union and complement) (cf.\ \cite{Davey:2002}). For the lattice of subspaces, note that the disjunction $p \vee q$ does not correspond to the set union of the subspaces, but the span. In fact, the union of subspaces is not part of the lattice.  This is why the lattice, even if it is a complemented $0-1$ lattice, fails to be a Boolean algebra: it does not use the standard set operations.

Clearly, $\sigma$-algebras are Boolean algebras because they are exactly an algebra of sets with standard set operations. As we said before, the Borel algebra will contain the lattice of subspaces as well as all the set operations between them, including the union. In a way, the lattice of subspaces is not Boolean because it does not contain enough sets, which are instead contained in the Borel algebra.

At this point, one may ask: do these extra sets correspond to physically meaningful propositions? To answer this question, let us first demonstrate the following proposition.

\begin{prop}
	Let $\mathcal{H}$ be a Hilbert space and $\langle \, , \, \rangle$ be its inner product. Let $A$ be a self-adjoint bounded linear operator. Let $F_A : \mathcal{H} \to \mathbb{R}$ be the map defined by $F_A(\psi) = \langle \psi , A \psi \rangle$. Let $U \subseteq \mathbb{R}$ be a Borel set. Then $F_A^{-1}(U)$ is a Borel set.
\end{prop}

Proof. Let $A$ be a self-adjoint bounded linear operator. Recall that a linear operator is continuous if and only if it is bounded. Therefore $A$ is continuous. Note that $F_A$ is the composition of $A$ with the inner product. Recall that the inner product is a continuous function. Since $F_A$ is the composition of continuous functions it is also continuous. Recall that all continuous functions are also Borel measurable, therefore $F_A$ is Borel measurable. Let $U \subseteq \mathbb{R}$ be a Borel set. Since $F_A$ is Borel measurable, the reverse image $F_A^{-1}(U)$ is also a Borel set. This concludes the proof.

Since $\langle \psi , A \psi \rangle$ expresses the expected value for operator $A$, the proposition shows that the Borel algebra can express statements about the expectation value of observables,\footnote{Technically, position and momentum are not bounded operators. However, one can argue that measured position and momentum are indeed bounded as the acceptance of a detector is always limited. Note that all operators for which the spectral theorem applies can be written as the limit of a sequence of bounded operators.} such as ``the expected position $\langle x \rangle$ of the particle is between $x_0$ and $x_1$.'' This would correspond to the Borel set $\{ \psi \in \mathcal{H} \, | \, \langle \psi , A \psi \rangle \in (x_0, x_1) \}$.\footnote{In most cases, one assumes states to be normalized, which is what we have done here for simplicity. Renormalization would lead to the set $\{ \psi \in \mathcal{H} \, | \, \langle \psi , A \psi \rangle / \langle \psi , \psi \rangle \in (x_0, x_1) \}$, which is still a Borel set since renormalization is a continuous function over its domain $\mathcal{H} \setminus \{ 0 \}$.} Consider, for example, ``the expectation of $z$-spin is zero''. This would correspond to the set $V \subseteq \mathcal{H}$ that includes all states, and only the states, for which the expection of $S_z$ is zero. By Proposition 2, $V$ is a Borel set. In fact, the set $U = {0}$ is a Borel set of $\mathbb{R}$ and therefore $V = F_{S_z}^{-1}(U)$ is a Borel set of $\mathcal{H}$. However, $V$ is not a closed subspace. The set $V$ is not the full set $\mathcal{H}$ since, for example, the state with $z$-spin up is not in the set. However, $x$-spin up and $x$-spin down will be in the set $V$, and their span is the full space $\mathcal{H}$. Therefore $V$ is not a closed subspace. Our proposition, despite being physically interesting, does not correspond to a closed subspace but does correspond to a Borel set.\footnote{Note that we are not claiming that all Borel sets can be constructed in this way or that all Borel sets correspond to physically interesting propositions. First of all, it is not true: $\{ \psi \}$ and $\{ e^{\imath \theta} \psi\}$ are different Borel sets, yet they are not physically distinguishable since a difference in absolute phase is not physically relevant. This can be fixed by considering the Borel sets of the projective space. Second, to really examine the physicality of all Borel sets we need a ``general theory of experimental logic'', of the type provided by \cite{Kelly:1996} or \cite{Carcassi:2021}, which would go well beyond the present discussion. With appropriate caveats (e.g. the Hilbert space must be separable), one can say that the Borel sets correspond to propositions that can be associated to an experimental test (what \cite{Carcassi:2021} call ``theoretical statements'') which may or may not terminate in any or all cases. We leave this discussion for another work.}

The idea is that closed linear subspaces correspond to single-shot measurements. While these form an interesting subset of possible measurements, they are fairly limited in what they can tell us about the incoming system. For example, measuring $z$-spin up tells us very little about spin \textbf{before} the measurement: only that it couldn't be $z$-spin down. To make any inference on the incoming state we need to gather statistics. In fact, no cross section measurement, our main tool in exploring the standard model, is single-shot. Being that quantum mechanics is a statistical theory, most physically interesting statements are statistical in nature, and that is why we need to include the Borel sets in our lattice of statements.\footnote{Since probabilities are assigned to Borel sets and only Borel sets, the Borel algebra will contain all physically interesting statements. Further expansions (e.g. to the power set) would only include unphysical statements.}

Another distinction is important. Borel sets allow us to express statements such as ``the expected position $\langle x \rangle$ of the particle is exactly $x_0$''. This is \emph{not} equivalent to the statement ``the particle is in the eigenstate of position corresponding to $x_0$'' This latter proposition will be true only if the whole function is at $x_0$. That is, if $x_0$ is the support of the wave function. The former is much less stringent: the wave function can even be distributed over an infinite range (e.g.\ a Gaussian wave packet) just as long as the average is $x_0$. This means that we can have statements like ``the expected position $\langle x \rangle$ of the particle is exactly $x_0$''$\wedge$ ``the expected momentum $\langle p \rangle$ of the particle is exactly $p_0$'' without having contradictions (e.g.\ a Gaussian wave packet can be centered on any value of position and momentum).

We can go one step further. Note that all statements that characterize the position and momentum of the center of mass form a sub-algebra of the Borel algebra. Recall that the Ehrenfest theorem states that $m \frac{d}{dt}\langle x \rangle = \langle p \rangle$ and $ \frac{d}{dt}\langle p \rangle = - \langle V(x) \rangle$. If the potential for a quantum particle can be considered constant over the wave function (i.e. $\langle V(x) \rangle = V(\langle x \rangle)$ ), the motion of the center of mass will reduce to the case of a point-particle. The Borel algebra contains as a sub-algebra the statements we need for (at least one form of) the classical limit. And this is a purely classical sub-algebra, because the average position and average momentum can both be well defined. Moreover, note that this is not dissimilar to what happens in classical mechanics when, for example, we study planetary motion using only the position and momentum of the center of mass. The algebra of statements for the position and momentum of the center of mass can be independently described and studied. In this regard, the logical structure of the Borel algebra of quantum mechanics works in the same way.

\subsection{The Non-Distributive Lattice of Classical Mechanics}

Now that we have argued that quantum mechanics can be given an algebra that looks more like the one of classical mechanics, let us show that we can (at least formally) give classical mechanics an algebra that looks like the one of quantum logic. Given that quantum mechanics is inherently a probabilistic theory, let us compare it to a classical probabilistic theory. Thus, let us consider the space of all possible probability distributions $\rho(x,p)$ over phase-space. As already stated, we can formalize this as a Hilbert space, where a vector $\psi$ corresponds to $\sqrt{\rho(x,p)}$. This will be a real vector space, and we are only going to consider operators that can be expressed as a multiplicative function $f(x,p)$. Note that $\langle \psi | F \psi \rangle = \int f(x,p) \rho(x,p) dx dp$ as one would expect.\footnote{To be clear, we are not arguing that this construction is the most appropriate or useful. Simply that it can be done.}  We can now take the algebra of subspaces and, as in the quantum case, it will be non-distributive: the disjunction of two subspaces will not be the set union, but the span.

To really understand the source of the non-distributivity, let us study the simplest possible example. Let us assume that there are only two possible classical states of a certain physical system $s$, $1$ and $2$. A state is identified by a two-valued distribution $[\rho_1, \rho_2]$, where $\rho_1$ is the probability to find the object in state 1 and $\rho_2$ is the probability to find the object in state 2. We denote with $[k,0]$ the subspace spanned by the first state only. This corresponds to the statement ``the system $s$ is certainly in state $1$'' since the probability for state $2$ is zero. Conversely, $[0,k]$ corresponds to the statement ``the system $s$ is certainly in state $2$''. The disjunction $[k,0] \vee [0,k]$ will correspond to the subspace spanned by both, which is the whole space $[k,j]$ where we can assign different non-zero probability to either case. We can also take the subspace $[k,k]$, the subspace spanned by vectors that have equal components in $1$ and $2$, which corresponds to the statement ``the system is equally likely to be in either state $1$ or $2$''. Note that $[k,0]$, $[0,k]$ and $[k,k]$ are all pairwise disjoint: no pairs have a vector in common apart from zero. Therefore their conjunctions $[k,0] \wedge [0,k] = [0,k] \wedge [k,k] = [k,k] \wedge [k,0] = [0,0]$ are zero dimensional. On the other hand, $[k,k]$ is a subspace of the whole space $[k,j]$ and therefore $[k,k] \wedge [k,j] = [k,k]$. Putting it all together, we have $[k,k] \wedge ( [k,0] \vee [0,k] ) = [k,k] \wedge [k,j] = [k,k]$. But we also have $( [k,k] \wedge [k,0] ) \vee ( [k,k] \wedge [0,k] ) = [0,0] \vee [0,0] = [0,0]$. Therefore the lattice is not distributive.

Note that during the discussion there was no mention as to whether $k$ and $j$ were real numbers, complex numbers or quaternions. The argument is formally the same in all cases. Therefore the lack of distributivity has nothing to do with the complexity of the Hilbert space or non-commuting observables. It's a property of the lattice of subspaces of any vector space and, more generally, of the lattice of subgroups of any group. Indeed, as a further example, consider the group of boost transformations in relativity theory. The boosts along one direction form a subgroup. If we take the subgroups of boost along $x$ and $y$ respectively, the smallest subgroup that contain both, the disjunction, is the subgroup of all boosts along any direction in the $x$-$y$ plane. As before, the disjunction is the span, hence the resulting set is different from the union, and the union of all boosts along the $x$-axis and the $y$-axis is not a subgroup and therefore not contained in the lattice. Thus, this has the same non-distributive structure as quantum logic.

We want to stress once again that we are not claiming that the lattice of subspaces of quantum logic is not of interest, or that it is of equal interest in both classical and quantum mechanics. Here we have pointed out that the argument that quantum mechanics \emph{implies} a radical departure from classical logic is untenable, given that both theories are built on similar algebraic structures. We conclude that such claims, still believed to this day, should be substantially weakened, in the light of the arguments that we proposed in this section. 

\section{Concluding Remarks}
\label{conc}

In this essay we considered the question of whether QM entails a rejection of classical logic. After having reviewed the main arguments usually given in literature to support such a claim, and having introduced the standard approach to quantum logic, we showed not only that the classical distributivity law can be maintained in the context of quantum theory, but also that the alleged peculiar behavior of quantum disjunctions---generally ascribed to the existence of superpositions---is easily explained once the meaning of quantum propositions is unambiguously defined. 

More precisely, in this essay we argued that in order to understand the actual meaning of quantum propositions it is important to take into account whether they refer to preparations or measurements. Indeed, it is the temporal ambiguity present in the example discussed in Section \ref{QL} that led to the wrong conclusion according to which the distributive law is refuted by the physical content of standard quantum mechanics. Contrary to this claim, we argued that making explicit the temporal dependence of logical propositions is sufficient to recover \eqref{DL} in the quantum context. In addition, this result entails that the non-commutativity of quantum observables does not play any role in refuting or rejecting the distributivity law, meaning that this peculiar algebraic property of quantum operators does not change the rules of logic, dissolving one of the most important myths of QL. Similarly, we explained that the existence of superpositions does not entail per se that quantum disjunctions are true even in the case in which neither of its members is. 

Furthermore, in Section \ref{Math} it has been shown that taking into account the relevant and fundamental algebraic structures on which quantum mechanics is built, i.e.\ the Borel algebra, it is possible to demonstrate that quantum theory can generate a distributive lattice. Moreover, we argued that considering the space of probability distributions over phase space one can assign a non-distributive lattice to classical propositions. These facts indicate that neither the complexity of the Hilbert space characterizing quantum theory nor the non-commutativity of quantum observables entails any revision of the rules of classical logic in the quantum domain. Referring to this, going back to the very mathematical foundations of both classical and quantum theories, we saw that the Borel algebra, which is a Boolean algebra, characterizes the formal edifice on which both theories are built, and guarantees that both frameworks can be given a distributive lattice of logical propositions, showing as a consequence a remarkable similarity. The non-distributivity of quantum lattices originates from the simple algebraic fact that they do not employ standard set operations. However, this feature is a property of the lattice of subspaces of \emph{any} vector space, and of the lattice of subgroups of any group more generally. Hence, we can safely conclude that the mathematical structures on which quantum theory rest do not entail a rejection of classical logic. 

Finally, let us conclude by saying that once we understand the similarity between the theories, we are in a much better position to understand their differences, which naturally exist. For example, quantum mechanics can only be expressed as a probabilistic theory and therefore \emph{always} requires a structure that in classical mechanics is not mandatory. In quantum mechanics, all subspaces of dimension one represent a pure state while in classical mechanics this is not the case. If we take two intervals $U_x$ and $U_p$ in position and momentum, and form the statement ``the state is wholly in $U_x$ and $U_p$'', meaning that the vector components are zero outside of those regions, in the classical case we always find at least one $\rho(x,p)$ that satisfies the bounds, while in quantum mechanics, if the bounds violate the uncertainty principle, no suitable $\psi(x)$ can be found. The point at stake in this paper---which is often not adequately recognized---is that the differences  are in the content of the logical structure of these theories, i.e.\ on the particular relationships that exist among the logical statements, and not on the type of structure.
\vspace{5mm}

\textbf{Acknowledgements:} withheld from draft.
\clearpage

\bibliographystyle{apalike}
\bibliography{bibliography}
\end{document}